# Compressive deformation of Fe nanopillar : Modalities of dislocation dynamics

Amlan Dutta

S. N. Bose National Centre for Basic Sciences, Block – JD, Sector – III, Salt Lake, Kolkata, India – 700 106

**ABSTRACT**

Unlike the tensile mode, compressive deformation of a bcc metallic nanostructure is mediated by the glide of screw dislocation. Although the bcc screw dislocations are well known to possess unusual attributes, it is still unclear how these unique effects manifest in a nanoscale solid. In the present study, atomistic simulations render a close look at the dislocation activities underlying the compressive deformation of bcc iron nanopillars. It is found that instead of performing simple glide motion, the line defects exhibit a host of complex features. In this regard, the temperature is observed to have a pronounced effect on the dislocation mechanisms and consequently, on the overall plastic response of the material. Additionally, statistical features of the load-strain data have been explored.

Email: amlan.dutta@bose.res.in

# 1 Introduction

Over the last several years, a great deal of effort has been made to understand the mechanical behavior of ultrathin one-dimensional structures like nanopillars and nanowires. As a result, numerous phenomena of interest have been discovered with regard to the plastic deformation of such nanostructures. Interestingly, there is a lack of generality and every system is associated with a set of features unique to the type of the material. For instance, ceramic nanostructures show peculiar stress-strain effects due to factors like phase transformation [1] and microstructural variations [2]. Similarly, the method of processing is known to dictate the mechanical properties of Cu-Zr metallic glass nanowires [3]. In the case of crystalline metallic one-dimensional nanostructures, size-dependent mechanisms govern the plasticity in fcc Au nanowires [4], whereas the hcp Co nanostructures are linked to the phenomenon of pseudo-elasticity at large strain [5].

Keeping up with the general trend, the bcc metallic nanostructures also have their fair share of uniqueness. The most elaborately studied effect in such materials is the compression-tension asymmetry. In general, a one-dimensional metallic nanostructure is expected to show qualitatively similar trend of deformation regardless of whether the deformation is tensile or compressive. This is because in both cases the magnitude of resolved shear stress on the glide plane remains the same. However, experimental investigations on bcc nanostructures have revealed fundamental differences between the deformation behaviors under tensile and compressive loadings. The stress vs. strain plots obtained during the compressive deformation of bcc metallic nanopillars display characteristic serrations, a feature clearly absent in the tensile deformation [6]. Despite several experimental studies [6-11] carried out to observe the plastic deformation of bcc nanopillars, a detailed understanding of the underlying mechanisms could not be obtained. In this backdrop, atomistic simulations have played a key role in unveiling the mechanistic details of these deformation processes. Molecular dynamics (MD) simulations performed by Healy and Ackland [12] have shown different deformation mechanisms at work during the tensile and compressive loading of a bcc iron nanopillar. The tensile deformation is found to occur through twinning, whereas the glide of screw dislocations is deemed responsible for the compressive deformation. This difference was initially attributed to the twinning-antitwinning asymmetry [6]. Subsequently, the three-dimensional spreading of bcc screw dislocation core has also been proposed as a significant factor contributing to the compression-tension asymmetry [12].

Although most of the experimental investigations have dealt with the compressive deformation of bcc nanopillars, simulation-based studies have followed the opposite trend. Majority of the atomistic simulations have aimed at understanding either the compression-tension asymmetry or specifically the mechanism of tensile deformation. In contrast, there has been little interest in the deformation under compressive load. This is surprising in view of the fact that screw dislocations are known to mediate the plasticity during compressive deformation. Screw dislocations in bcc crystals are well known as special line defects on account of their astonishing characteristics. Firstly, the response of a bcc screw dislocation to an applied shear load is known to depend upon its core structure, which can be polarized or un-polarized [13]. Secondly, the behavior of a bcc screw can be sensitive to the non-shear component of the stress tensor [14]. Thirdly, its dynamics exhibits smooth-to-rough and rough-to-twinning transitions depending upon the applied load and temperature [15]. Because of these peculiar effects, it is reasonable to expect non-trivial features in the mechanism underlying the compressive deformation of bcc nanopillars. Clearly, there is a need to take a closer look at how these nanostructures respond under compressive loading.

In view of the above arguments, the present study examines the compressive deformation of bcc-iron nanopillar through MD simulations with particular emphasis on the modalities of dislocation motion. To this end, advanced algorithms for dislocation-extraction and pattern-matching have been used to characterize the defects in the nanostructure. The simulation outcome reveals several fascinating dislocation phenomena accommodated within the small domain of the nanopillar. These include dislocation reactions, formation of point defect debris and nucleation of twin-faults. Furthermore, the statistics of load serrations obtained from these virtual, *in-silico* experiments is analyzed and compared to those obtained from theory and real experiments.

## 2 Scheme of Simulation

The present study follows a simulation strategy similar to the one used by Healy and Ackland [12], except for a smaller strain-rate. The bcc-Fe nanopillar sample is shown in the inset of Fig. 1(a). The <100>-oriented pillar has 5.8 nm wide square cross-section and height of 15.4 nm. The side facets are {110}-surfaces with low energy. The nanopillar is supported between two rigid plates as shown in the figure and the forces on these plates are measured to compute the load generated during the deformation. In order to model the bcc phase of iron, the potential developed by Mendelev *et al*. [16, 17] has been employed. Apart from reproducing the basic material properties, this potential exhibits the non-degenerate core-structure of screw dislocation [13] in consonance with DFT calculations [18]. The system is thermalized at the desired temperature before initiating the compression. Subsequently, a compressive strain rate of $10^8$ s$^{-1}$ is imposed by gradually shrinking the dimension of the simulation cell in axial direction while maintaining a constant temperature using the Berendsen thermostat [19]. Deformation is carried out at 50 K, 300 K and 500 K to cover a broad range of temperature. For each of the three temperatures, set of fifteen simulation runs were performed with different initializations of atomic velocities. Simulation snapshots and other data presented in the following section are representative examples, which cover the basic features of simulation outcome. All the simulations reported here are executed with the MD code LAMMPS [20]. A newly developed dislocation extraction method and pattern matching algorithm (implemented through the Crystal Analysis Tool) serve the purpose of identifying the crystal defects [21-23]. The OVITO package [24] is used for graphical visualization and other minor analyses.

## 3 Results and discussion

### 3.1 Effect of temperature on yield stress

In all the simulations discussed here, the nanopillars are compressed to a sizeable strain of 30%, during which the cross-sectional area of the pillar varies along its length. Therefore, instead of presenting the stress on the nanopillar, the results are expressed directly in terms of the measured forces on indenter plates for the purpose of general and unambiguous representation. Figure 1(a) shows the load *vs*. strain plots at the three temperatures, where the jerky flow exactly resembles the trend witnessed in Ref. [12]. Before the onset of plasticity, a non-Hookean behavior can be observed at large elastic strain, which is most prominently visible in the 50 K simulation. Moreover, we find that the strain and load at the yield point increases with reduction in temperature. Table 1 shows that reducing the temperature from 500 K to 300 K causes a 17.1% increment in the yield stress and 20.5% increment in the yield strain. These increments become 62.2% and 61.6%, respectively, as the temperature is brought down to 50 K. Even without probing the comprehensive mechanism of deformation, we can perceive that the thermally activated process of nucleation of dislocation [25] renders the yield point temperature-sensitive.

To further elaborate the effect of temperature, it is helpful to take a look at the previous studies reporting on the compressive deformation of such nanostructures. In contrast to several simulation-based investigations on the tensile deformation of one-dimensional bcc nanomaterials, only two detailed studies have dealt with the case of compressive deformation. In the work of Healy and Ackland, the stress-strain plot (Fig. 2(a) in Ref. [12]) indicates a jerky deformation with intermittent stress-drops of significant magnitudes, which is qualitatively analogous to the serrated yielding found in the experiments [6]. On the other hand, Sainath and Choudhary presented a different result (cf. Fig. 3 in Ref. [26]), where the elastic limit was found to extend to a very large compressive strain of about 20% and a stress of 30 GPa. This yield stress is not only a few times larger than what is observed in Ref. [12], but also excessively large in comparison to the experimental findings [6, 7, 10]. Although both studies recognize the dislocation-mediated plasticity as the dictating mechanism, a comparison of their fundamental methodologies is necessary to understand the possible reasons of this discrepancy. While Healy and Ackland [12] simulated nanopillar with <100> orientation and {110} lateral faces, the <100> nanowire of Sainath and Choudhary [26] had {100} facets with higher surface energy. Moreover, the nanowire had pseudo-continuity on account of the periodic boundary condition in axial direction. By contrast, the nanopillar setup in Ref. [12] had true volume confinement due to rigid indenter plates at both ends of the pillar. Apart from these methodological differences, the most conspicuous aspect is the temperature maintained during deformation. While the nanopillar was deformed at room temperature, the nanowire was compressed at an exceptionally low temperature of 10 K [26]. Observing the trend of temperature-effect as observed in Fig. 1(a), it is not surprising to find the extraordinarily large stress and strain at yield-point in the 10 K deformation of bcc-Fe nanowire. As a matter of fact, this magnitude of stress may be unrealistic from a pragmatic standpoint, for a real nanostructure with large aspect ratio would rather buckle due to elastic instability [27, 28], than undergo an axial compressive deformation. The computed stresses seen in Fig. 1(a), however, are much smaller and match well to the order obtained from the experiments.

### 3.2 Elementary dislocation processes

A common behavior is found in the three plots in Fig. 1(a). In each case, the slow build-up of load is punctuated by sudden rapid drop, which is suggestive of intermittent bursts in dislocation activities. To have a better perspective of such events, the evolution of defect structure of the nanopillar is closely monitored. Due to stress-concentration and presence of free surfaces, corners of the pillar act as preferred sites for the nucleation of dislocations, which are typically of Burgers vector <111>/2. Initially the line defects are curved and the line direction continuously changes from predominantly edge to screw orientation along the line of the dislocation. With increase in load, the edge component tends to glide much faster due to larger mobility, thereby quickly straightening the originally curved line to full screw orientation as demonstrated in Fig. 1(b). The first glide activity marks the yield point of the pillar. Nevertheless, careful examination reveals that the simple glide motion of dislocations do not solely dictate the compressive deformation. After the yield point is attained, several non-trivial defect structures are observed inside the material. These include – (1) <100> and <110>-dislocations, (2) point defects and (3) twin plates. On the whole, compressive deformation of the pillar involves mechanisms associated with these point, line and planar defects, which are often acting simultaneously. As discussed in the subsequent subsections, even the point and planer defects are associated with the dynamics of dislocations. It is thus advantageous to resolve the complex process of plastic deformation into elementary dislocation activities. This understanding will further aid us in assimilating the mechanism of compressive deformation in a holistic manner, where the different elementary processes overlap and interact to produce the gross deformation behavior of the Fe nanopillar.

### 3.2.1 Formation of <100> and <110>-dislocations

In addition to the primary dislocations of type <111>/2, secondary dislocations with other Burgers vectors are also found in bcc metals. In particular, the presence of <100>-dislocations has been a well-established observation since decades [29]. It is generally assumed to form by the intersection of two <111>/2-dislocations. In the present study, secondary dislocations are found to form inside the deforming nanopillar. One such example is shown in Fig. 2(a), where two <111>/2-dislocations are seen to attract each other. They merge to form an X-junction and subsequently, a zipping process creates a segment of <100>-dislocation according to a reaction of the type,

$$\tfrac{1}{2}[111]+\tfrac{1}{2}[1\bar{1}\bar{1}] \rightarrow [100]. \quad (1)$$

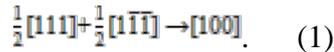

The <100>-segment contains two Y-junctions at its ends. As demonstrated in Fig. 2(a), the zipping can continue, thereby creating a standalone <100>-dislocation. However, the most common observation is that of short <100>-type segments, which form parts of a larger dislocation network (cf. Fig. 2(b)).

Another type of line defect is the <110> dislocation, which can emerge from the reaction,

$$\tfrac{1}{2}[111]+\tfrac{1}{2}[11\bar{1}] \rightarrow [110]. \quad (2)$$

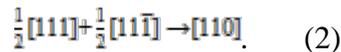

This reaction, despite satisfying the continuity of Burgers vector, is never considered due to the energetic unviability arising from Frank's criterion ($\mathbf{b_1}\cdot\mathbf{b_2} < 0$ for favorable reaction). Nonetheless, such <110> segments can form at an X-junction (Fig. 2(c)) under the effect of stress and thermal fluctuations. Due to thermodynamic instability, the above reaction is transient in nature. The <110>-segments are always very short and quickly revert back to the original X-junction configuration.

### 3.2.2 Formation of point defects

During the course of deformation, a large number of point defects are produced inside the nanopillar. Unlike the conventional thermodynamic formation of point defects, these are created by the actions of gliding screw dislocations. The typical formation of deformation-induced vacancies is exhibited in Fig. 3, where we see that the passage of a screw dislocation increases the concentration of vacancies in the pillar. Such vacancies constitute the debris, which is an evidence of the so called 'roughening' of dislocation glide. A detailed explanation of this phenomenon was offered by Marian *et al*. [15], who proved that these point defects resulted from the process of cross-kinking, where spatially separate parts of a screw dislocation line undergo kink-pair nucleation on different but equivalent {110} planes. Consequently, as the kink-pairs expand and try to merge, they fail to recombine and a conflict arises. This causes a strong pinning of the gliding dislocation, which is overcome through the creation of small self-loops on the dislocation line. In the present simulations of nanopillars, the self-loop is often so small than a single point-defect is created; however, divacancies are also observed occasionally. Marian *et al*. reported that the transition of dislocation motion from smooth to rough dynamics was facilitated by thermal assistance [15]. In the strain-controlled deformation of bcc nanopillar, the reduction in temperature to 50 K entails large values of yield and flow stresses (Fig. 1(a)). This increment in stress compensates for the loss in thermal assistance (cf. Fig. 2 in Ref. [15]). In addition, image stress on the dislocation due to free surfaces of the pillar has been predicted to promote the process of self-looping [30]. This explains the observation of significant production of point-defect debris even in the low-temperature simulation. Due to much larger energy of formation, interstitials are observed less frequently and the deformation debris is mainly dominated by the vacancies.

### 3.2.3 Formation of twin faults

In addition to exhibiting a smooth-to-rough transition associated with the self-pinning phenomenon of screw dislocations, MD simulations by Marian *et al.* have also established a rough-to-twinning transition occurring at very large shear load [15]. In this transition, self-pinning caused by the point defect debris resists the glide of a screw dislocation, thereby disabling it from creating the high plastic strain rate commensurate with the large applied stress. Under such conditions, the dislocation nucleates a twin plate in order to maintain the large strain rate. The feasibility of this process was first suggested by Sleeswyk [31], who proposed the following dissociation of the <111>/2-screw dislocation into three twinning fractional dislocations,

$$\frac{1}{2}[111] \rightarrow \frac{1}{6}[111] + \frac{1}{6}[111] + \frac{1}{6}[111]. \quad (3)$$

Under the effect of resolved shear stress, these twinning dislocations were predicted move along the same direction on three parallel {112} planes to create a three-layer micro-twin [32]. Marian *et al.* [15] showed that once the micro-twin was nucleated, further nucleation and expansion of twinning dislocation-dipoles occurred on the back of the twin-boundaries. This caused the twinned region to grow in thickness.

Noticeably, the present simulations of nanopillar deformation also show the formation of twin nuclei. These are typically in the form of twin ribbons of small width. The simulations reported in Ref. [15] had a simple configuration with a single <111>/2-screw in the simulation cell. Therefore, self-pinning was evidently the only mechanism responsible for impeding the motion of the line defects. Nevertheless, a dislocation inside the deforming nanopillar can also get pinned by means of external factors like formation of junctions or elastic interactions with other dislocations. Thus, the dissociation of a screw dislocation into twin nucleus can occur even without self-pinning. An example of such dissociation is shown in Fig. 4(a), where an entire dislocation line dissociates into a narrow twin-ribbon bounded by twinning dislocations. Several variations of this process are observed in the simulations. For example, Fig. 4(b) shows a partial dissociation, where only a part of the dislocation is dissociated into a twin fault. Similarly, Fig. 4(c) shows a junction between two twin-ribbons, which is roughly analogous to the X-junction between two <111>-screw dislocations as displayed in Fig. 2(a). Besides, Fig. 4(d) shows that a twin ribbon can also interact with a full dislocation to form a hybrid ribbon-dislocation junction. Once a full dislocation splits into a twin-ribbon, its mobility is severely restricted. Therefore, networks like the one shown in Fig. 4(d) play a key role in determining the strength of the bcc nanopillar during compressive deformation. Once formed, a twin-ribbon often becomes part of a dislocation network and stays intact for a significant duration. It is eliminated only upon accidental interaction with an annihilating full dislocation with opposite Burgers vector.

Even though the simulations show the formation of twin faults within the pillar, a detailed analysis indicates a mechanism that is somewhat different from what was assumed in the previous studies [15, 31, 32]. This is evident from Fig. 5(a), which displays the atomic configuration in a part of the nanopillar just after the dissociation of a <111>/2-screw dislocation. It is interesting to note that the dissociation yields a four-layer twin fault, instead of the expected three-layer micro-twin. Although the two middle planes show typical relative slips of <111>/6, the two boundary layers have slips of <111>/12 each. Hence, instead of observing the conventional reflection twin-boundaries, we witness the displaced or isosceles twin-boundaries in these simulations. This suggests that the <111>/2-screw actually dissociates not into three, but four fractional dislocations according to the reaction,

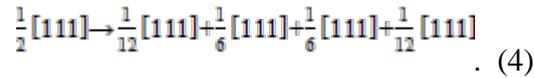
. (4)

Based upon the results of Bristowe *et al*. [33], the possibility of such dissociation was first conjectured by Christian [34]. Moreover, the resulting twin plate has a structure very close to that of the stable four-layer stacking fault predicted by the computations of Machová *et al*. [35]. Another interesting aspect is the finding that soon after its formation, the four-layer twin ribbon grows into a six-layer ribbon shown in Fig. 5(b). This growth can be interpreted as the formation of two <111>/6 dislocation-dipoles, one on each side of the four-layer twin structure. Each of these <111>/6 twinning dipoles can further split into two <111>/12 dipoles on two adjacent twinning planes and therefore, displaced boundaries are always found even in a twin structure which has grown to six layers as shown in Fig. 5(b). This mechanism is similar to the recently discovered modality of twinning in tensile deformation of bcc materials [36]. It must be noted that a large stress of the order of GPa is typically required to dissociate a screw dislocation into partials. In comparison, a twinning dipole formed on the back of a pre-existing twin boundary experiences a very small back-force. Therefore, if the local stress is large enough to cause dissociation of a screw dislocation into a four-layer twin, it also suffices for its thickening as well. This has been a basic premise of Sleeswyk's analytical work [31].

Although the self-pinning at large shear load is known to cause the dissociation of a screw dislocation into a micro-twin, dislocation pinning at a pre-existing vacancy can also trigger similar dissociation. For instance, Fig. 6 shows a moving <111>/2-screw dislocation (D) heading towards the free surface (S) of the nanopillar. In this projection, two vacancies ($V_1$ and $V_2$) can also be seen, which were created earlier by other dislocations passing through this region. The gliding screw impinges upon one of the vacancies ($V_2$), which strongly arrests its motion. This triggers its immediate dissociation into a twin nucleus, which extends towards the surface and gets terminated. Hence, the large concentration of point-defect debris is not only associated with the intrinsic transitions of dislocation dynamics through self-pinning, but may also cause such transitions extrinsically.

### 3.3 Mechanism of compressive deformation

Having gathered a comprehensive insight into the basic dislocation activities occurring inside the deforming pillar, it is now possible to comprehend the exact mechanism by which the nanopillars deform. As the load-strain behavior is sensitive to temperature, the process of deformation should be considered separately for the temperatures of 50 K, 300 K and 500 K.

### 3.3.1 Deformation at 50 K

Figure 7 displays the simulation snapshots of the Fe nanopillar undergoing compression at 50 K temperature. Due to lack of thermal assistance, nucleation of dislocations is deferred until a large compressive strain (about 11.7%) is attained. During the motion of screw dislocations, a large number of point defects are created through the mechanism of cross-kinking as described in Sec. 3.2.2. The dislocations frequently intersect to create networks with junctions and dislocation segments with Burgers vector <100> (cf. Sec. 3.2.1). As shown in the figure, dissociation of <111>/2-screw into twin-ribbon (cf. Sec. 3.2.3) is also a common phenomenon at this temperature and multiple ribbons may be present simultaneously inside the pillar. In this example, a dislocation is seen to transform into a twin nucleus at a strain of about 26%. Nevertheless, unlike the previously formed twin-ribbons of small width, it rapidly expands laterally along the {112} planes. At the same time, it also thickens vertically due to repeated nucleation and expansion of twinning dislocation dipoles, thereby increasing the distance between the twin-boundaries. The thick twin slab thus created has one end inside the nanopillar, while the other end terminates at the surface as shown in the last snapshot of Fig. 7.

Although the simulation snapshots capture the essential components of the deformation process, a better perspective can be gained by means of the supplementary video S1 (in the ESM). The video exhibits the evolution of crystal defects in the Fe nanopillar alongside the load *vs.* strain plot in dynamic synchronization. We can clearly observe a direct relation between the sudden load-drops and dislocation activities. The simulation indicates that the glide of dislocation is not a continuous activity and occurs in bursts. Every such burst coincides with a load-drop in the accompanying load-strain plot. In between two consecutive bursts, dislocations remain immobile in locked configurations while the load keeps increasing. In the given example of deformation at 50 K, the growth of the twin slab can be seen to cause a large load relaxation. In general, a small drop in load is typically associated with the glide of a single line defect, whereas the collective dynamics of multiple dislocations are responsible for load-drops of larger magnitudes.

### 3.3.2 Deformation at 300 K

At room temperature, thermal activation facilitates the nucleation and glide of screw dislocations. Here the microstructure is dominated by screw dislocations. However, twinning can still occur, albeit with much reduced frequency. A particularly interesting instance of deformation at 300 K is illustrated in Fig. 8, where a twin fault is nucleated after a vacancy pins down a moving <111>/2-screw dislocation. This nucleation is the same event which has been detailed in Fig. 6. Once the nucleus is formed, it grows in both thickness and width to create a twin slab. The lateral expansion of this slab is complete, for it spans from one surface of the nanopillar to the opposite surface. The supplementary video S2 (in the ESM) renders a dynamic view of this process. A general aspect of the twinning process is highlighted in the inset of the last snapshot (30% strain) in Fig. 8. It shows a lateral, cross-sectional slice containing part of the twin slab. We can notice that just like the twin-ribbons (Fig. 5), larger twin-structures also have displaced twin boundaries corresponding to slips of <111>/12. The time evolution of this slice is presented in the supplementary video S3 (in the ESM), which clearly demonstrates the thickening process of the twin slab.

Unlike the 50 K deformation where large twin slabs are frequently formed, the example shown in Fig. 8 is an exceptional case which occurs only occasionally. The more common finding at 300 K is that of narrow twin-faults of small lateral widths. One such example obtained from a different simulation run at 300 K is demonstrated in Fig. 9(a), where we can see four separate twinned regions formed by the dissociation of four <111>/2-screws. We can see that unlike the fully formed slab displayed in Fig. 8, such small twins can dwell completely inside the crystal. As they laterally terminate within the material, they possess curved and closed boundaries at the sides. Sleeswyk [37] performed a detailed analysis of such non-coherent boundaries and concluded that the tip of a deformation twin of finite extent must be under extremely large shear load. He proposed a new mechanism to explain the stability of such twin-lamellae. According to this model, a <111>/6-type twinning dislocation bounding a twin fault can dissociate as,

$$\tfrac{1}{6}[\bar{1}\bar{1}1] \rightarrow \tfrac{1}{3}[111] + \tfrac{1}{2}[\bar{1}\bar{1}\bar{1}]. \qquad (5)$$

The <111>/3-dislocation, commonly known as the complementary twinning dislocation, replaces the dissociated <111>/6-dislocation at the border of the twin-fault and reduces the elastic energy of the structure. The <111>/2-dislocation resulting from the above dissociation is emitted and pushed away from the twin lamella. Although the original mathematical treatment [37] assumed edge dislocations, a similar process can also occur with the line defects of screw orientation. In fact, emissary screw dislocations have indeed been observed in the electron micrographs of ferrous alloys (cf. Fig. 6(b) in Ref. [38]).

The simulations suggest that the mechanism given in reaction (5) also plays a role in stabilizing the twin defects shown in Fig. 9(a). This is evident from the simulation snapshots given in Fig. 9(b), which show two emissary <111>/2-screws being expelled from the tip of the twin defect in quick succession. In Fig. 9(a), the two twin structures at the bottom left part of the nanopillar have actually formed by further dissociation of the two emissary dislocations shown in Fig. 9(b). In all likelihood, this is the first reported observation of emission of bcc <111>/2-dislocations by the tip of a twin lamella in direct MD simulation.

### 3.3.3 Deformation at 500 K

A qualitatively different scenario emerges during the compressive deformation at 500 K (Fig. 10). At such elevated temperature, the mobility of line-defects becomes very high [39] and they usually sweep through the nanopillar soon after they are nucleated. Besides, thermal assistance facilitates the depinning of dislocations despite the formation of a large number of point defect debris due to enhanced rate of cross-kinking. Therefore, the process involves negligible amount of dislocation storage and the pillar remains devoid of dislocations for most of the time. Due to increased screw mobility and absence of locked dislocation configurations, the imposed strain rate is entirely maintained by the gliding line defects without any need for formation of twin-structures. Occasionally, some twin nucleation may occur near the surface of the pillar; however, these nuclei are short-lived and rapidly exit the nanostructure. The supplementary video S4 (in the ESM) clearly highlights these features of compressive deformation at 500 K.

The effect of temperature on the density of screw dislocations is best understood by comparing the snapshots in Figs. 7 and 10. At 50 K, the deforming nanopillar hosts dense networks of screw dislocations, which can directly interact with the twin-boundaries as well. In contrast, the high mobility of dislocations causes a very low dislocation-density at 500 K. It can be seen that at 300 K temperature (Fig. 8), dislocation networks are created, albeit with smaller density as compared to those formed at 50 K. This is because of the moderate mobility of bcc screw dislocations at 300 K which is significantly larger than the 50 K mobility, but still smaller than that at 500 K [39]. Thus, the microstructure observed during the room temperature deformation can be considered as intermediate between the extreme cases of deformation at cryogenic and high temperatures.

### 3.4 Statistics of serrated yielding

Intermittent activities of dislocations are neither uncommon, nor exclusive to compressive deformation of nanopillars. Even in the bulk fcc and hcp crystals, acoustic-emission events are indicative of intermittent dislocation avalanches. Such events are known to exhibit scale-free statistics, which is reminiscent of self-organized criticality [40, 41]. As the system-size is reduced to nanoscale, jerky dynamics of deformation is directly evident from the serrated yielding of the material. We have already observed that the abrupt load-drops seen in Fig. 1 are associated with intermittent bursts in dislocation activities. A number of studies, both experimental and theoretical, have analyzed the statistics of such bursts. Three-dimensional discrete dislocation dynamics simulation of ultra-small volume of material under deformation has revealed that displacement-bursts are expected to follow a power-law statistics with 1.5 as the universal value of exponent [42]. This universal exponent is not only in agreement with the mean-field theory (MFT) [43], but has obtained experimental support as well. Brinckmann *et al*. [7] confirmed this power-law statistics for compressive deformation of both fcc and bcc nanopillars. However, this trend was found to be valid only for the bursts of large magnitudes. Less intense bursts, on the other hand, were found to have probability densities much smaller than those predicted by the mean-field power-law. Subsequently, a more refined and rigorous method of fitting has been able to show that that the actual exponent is smaller than the previously claimed value of 1.5 [9]. The recent numerical study by Ispánovity *et al.* [44] has revealed that a fundamental physical aspect lies at the core of such deviations from the predictions of MFT. The MFT is based upon the premise of pinning caused by quenched disorders in the medium [43]. Nevertheless, jerky dynamics of dislocations do not compulsorily need such static disorders. In particular, a clear distinction between pinning and jamming of line defects has been recognized, where the latter mechanism is observed to yield non-MFT exponents of power-law distributions [44]. More specifically, the exponents obtained from both continuous time dynamics and cellular automaton models are found to be very close to unity.

In the present work, magnitudes of intermittent load-drops are measured and their statistical distributions are displayed in Fig. 11. As already specified in Sec. 2, multiple simulations are carried out at each temperature. They reveal that the first drop in load after the yield point is always at the same strain, whereas the other load-drops occur randomly. This observation has also been reported by Healy and Ackland (cf. Fig. 3 in Ref. [12]). Therefore in order to avoid any non-stochastic component, the distributions presented in Fig. 11 ignore the first drop in each load-strain plot, and only consider the subsequent, stochastically significant values. Figures 11(a), (b) and (c) show the complementary cumulative distribution functions (CCDFs) of the load-drops ($\Delta F$) obtained at 50 K, 300 K and 500 K, respectively. It is interesting to find that in each case, a single statistical distribution cannot describe the whole range of load-drops obtained from the simulations. Instead, similar to the qualitative trend shown in Ref. [7], the regimes of small and large load-drops appear to follow different statistical trends. At all the three temperatures, the CCDFs in the regime of small drops follow the form, $\sim e^{-(\Delta F/\lambda)^{\alpha+1}}$, which suggests that the load-drops are Weibull-distributed as, $\sim \Delta F^{\alpha} e^{-(\Delta F/\lambda)^{\alpha+1}}$. Observation of Weibull-distributed measures in these simulations is not surprising. Experiments of load-controlled deformation of molybdenum nanopillars have revealed that while the strain bursts obey a power-law statistics, load-increment between two consecutive bursts exhibits the Weibull distribution [8]. The parameters $\alpha$ and $\lambda$ for these distributions are specified in Table 1.

In the regime of larger load-drops (cf. highlighted region in Fig. 11), the values are found to obey the power-law distributions with exponential cut-off ($\sim \Delta F^{\beta} e^{-\Delta F/\mu}$), in agreement with previous experimental [7] and theoretical studies [42]. However, instead of the MFT value of 1.5, a unique value of $\beta = 1$ has been obtained for all the temperatures (cf. Table 1). Although the simulations by Ispánovity *et al.* [44] have also predicted exponents close to unity, the fundamental differences between their 2-D models and the present MD simulations must be kept in mind. In Ref. [44], the statistics of strain avalanches were considered, whereas the load on indenter plate is the measured quantity in this study. Moreover, the present MD simulations involve a strong effect of dislocation reactions and formation of junctions. This effect, which is distinct from the jamming scenario arising out of the long-range elastic, has been recognized and thoroughly discussed by Zhang *et al.* [9]. In any case, the confirmation of power-law distribution and estimation of its parameters have been shown to be non-trivial tasks [45]. This proves to be even more difficult in this study for two reasons. Firstly, because of the physical nature of the observable, the datasets do not range over several orders of magnitude as typically seen in many other physical, biological and social systems. Secondly, as shown in Fig. 11, only a part of the whole range is available for fitting the power-law form. Therefore the idea of a possible relation between the observations in Ref. [44] and the present study, though intriguing, must be ascertained only through independent studies.

The MD simulations reported here can offer a qualitative perspective of the statistical trend seen in Fig. 11. We have already observed that the involvement of multiple dislocations is generally associated with load-drops of larger magnitudes, whereas the isolated movement of a single dislocation typically causes a small load-drop. Thus we can conceive the two regimes of a load-drop distribution as corresponding to two different categories of behavior of the line defects. The regime of large load-drop is dominated by collective dynamics, where dislocation segments get depinned at the junctions. By the way of contrast, we can consider the regime of smaller drops as unaffected by short range dislocation-dislocation interactions. It is therefore closer to the scenario explored by the two-dimensional atomistic simulations [46] where the mechanism of deformation does not involve dislocations-storage and is akin to the source-limited deformation. In such simulations, attempt to fit a power-law to the load-drops have been reported to yield significantly small value of the exponent. We can consider this observation as qualitatively similar to the general trend obtained in Fig. 11, where the slope in the regime of smaller load-drops is considerably less than that in the other regime.

## 4 Conclusions

This study employs the tool of atomistic simulation to unravel the intricacies of compressive deformation of bcc nanopillar. The underlying mechanism of deformation is found to depend upon the temperature of the material. A completely dislocation-mediated plasticity at high temperature turns into dislocation-plus-twinning mechanism as the temperature is reduced. Even when twinning occurs, it is found to proceed through the dislocation-dissociation mechanism, as distinct from the conventional twinning observed in tensile deformation. It is indeed fascinating to witness that a single simulation of such ultra-small volume can exhibit a host of features like dislocation reactions, production of point-defect debris, twin-nucleation and ejection of emissary dislocations. In addition, the simulations also attempt to explore the link between internal dynamics of line defects and statistical distribution of intermittent load-drops observed during the compressive plastic flow of the pillar.

Intriguing traits of plasticity, as disclosed by the computations, emerge with some concomitant questions that have been left out of the scope of the present work. One of these is the influence of image stresses on the various dislocation activities shown by the simulations. It will also be interesting to learn the possible effect of pre-existing dislocations inside the pillar is another curious issue. Similarly, the role played by the non-shear stress component is also a curious issue. Clearly, separate dedicated studies are needed to probe these aspects. It is expected that the results reported here will provide impetus for further investigations in these directions.

**Acknowledgements**

This work is supported under the INSPIRE faculty scheme by the Department of Science and Technology, Govt. of India. The author also thanks Dr. Alexander Stukowski for extending his help with the use of the Crystal Analysis Tool.

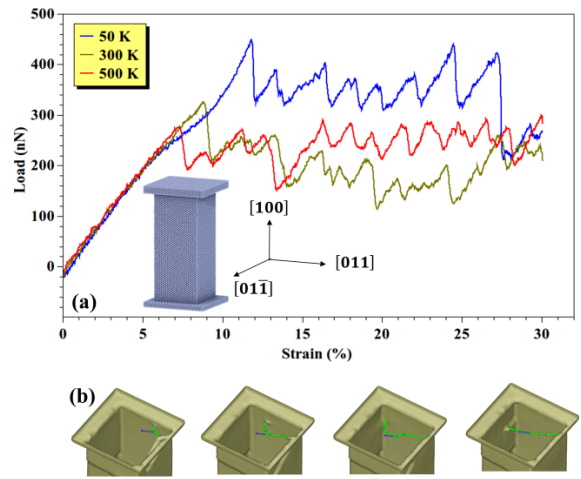

**Figure 1** (a) Load on the Fe nanopillar plotted against the compressive strain. Inset shows the simulated pillar with crystal directions. (b) Nucleation of a <111>/2-dislocation (green curve) at the corner of the pillar. The arrow is directed along the Burgers vector.

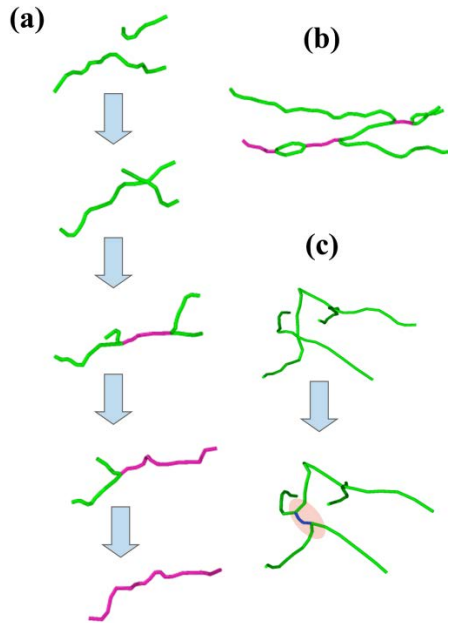

**Figure 2** Formation of secondary dislocations inside the nanopillar. Green, purple and blue lines represent line defects with Burgers vector <111>/2, <100> and <110> respectively. (a) Two <111>/2-dislocations merge to produce a <100>-dislocation. (b) A dislocation network containing short <100>-segments. (c) Formation of the transient <110>-segment at an X-junction.

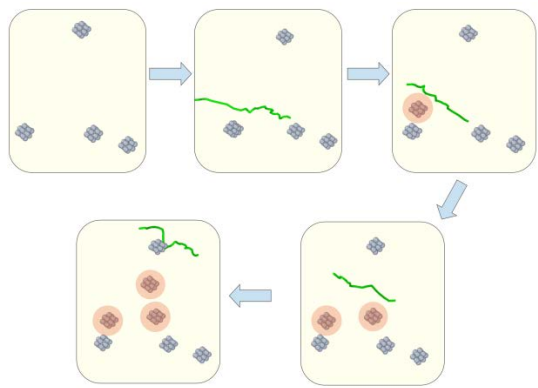

**Figure 3** Production of point defects by a gliding screw dislocation. The dislocation (green line) can be seen to create three new vacancies (highlighted) during its glide. The four pre-existing vacancies were created by other dislocations which passed through this region.

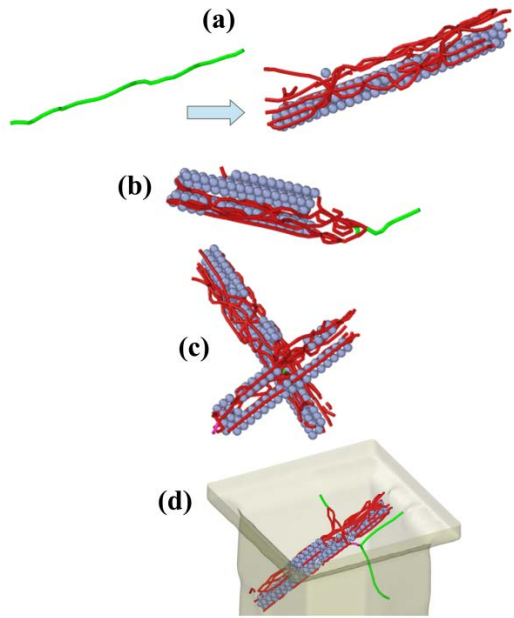

**Figure 4** Formation of twin ribbons through dissociation of screw dislocation. Spherical particles represent atoms at the twin boundaries identified by pattern matching, whereas the green lines indicate <111>/2-dislocations. Red lines are twinning dislocations surrounding the twinned regions. (a) A screw dislocation completely dissociated into a twin ribbon. (b) An example of partial dissociation. (c) Ribbon-ribbon junction formed by simultaneous dissociation of two screws. (d) Hybrid dislocation-ribbon network.

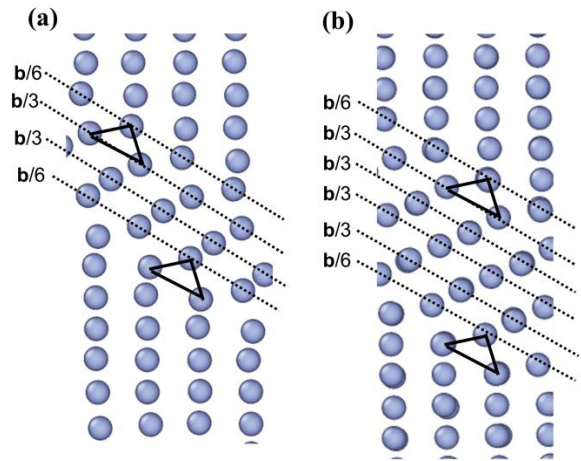

**Figure 5** A {110} atomic plane of the nanopillar showing the twin fault and relative slip of each layer. (a) Dissociation of a screw dislocation with Burgers vector **b** = <111>/2 produces a four-layer twin with displaced boundaries. It grows into a (b) six-layer twin due to subsequent nucleation of twinning dislocation dipoles.

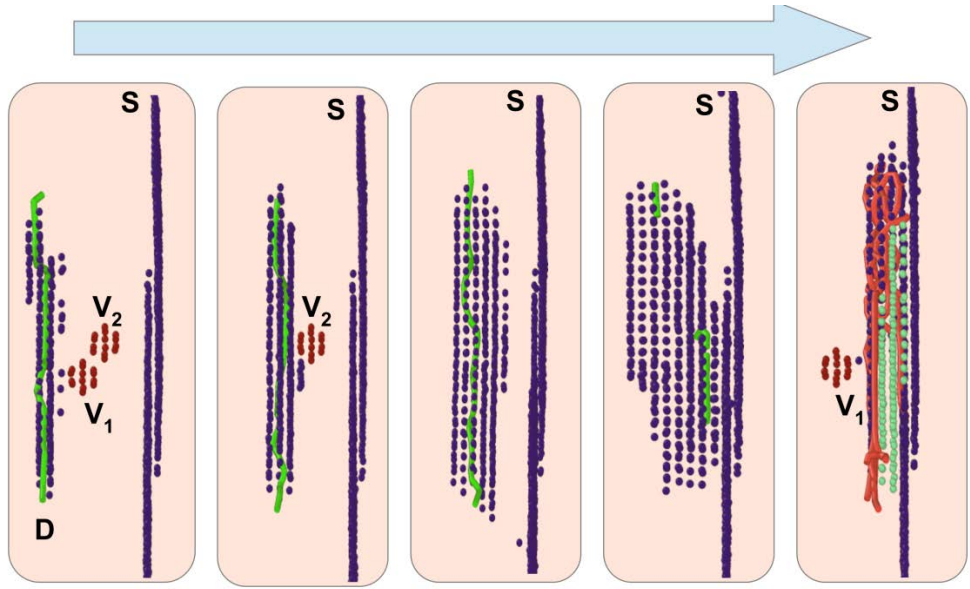

**Figure 6** Dissociation of a screw dislocation through extrinsic pinning at a vacancy. The twin-boundary atoms are shown in green.

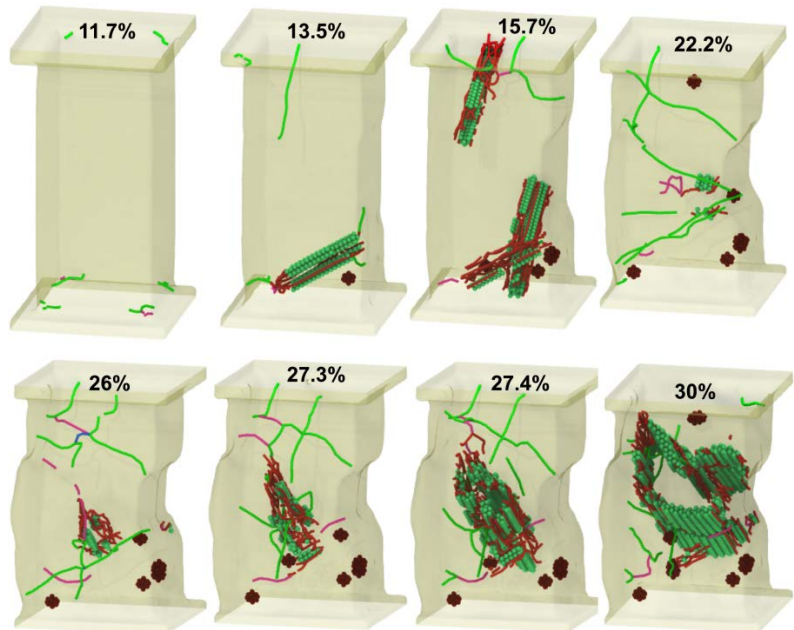

**Figure 7** Compression of the Fe nanopillar at 50 K. The simulation snapshots show various defects at different compressive strains (given in percentage). Atoms surrounding the vacancies are shown in dark red and those at the twin-boundaries are green. Various dislocations are shown as thin lines.

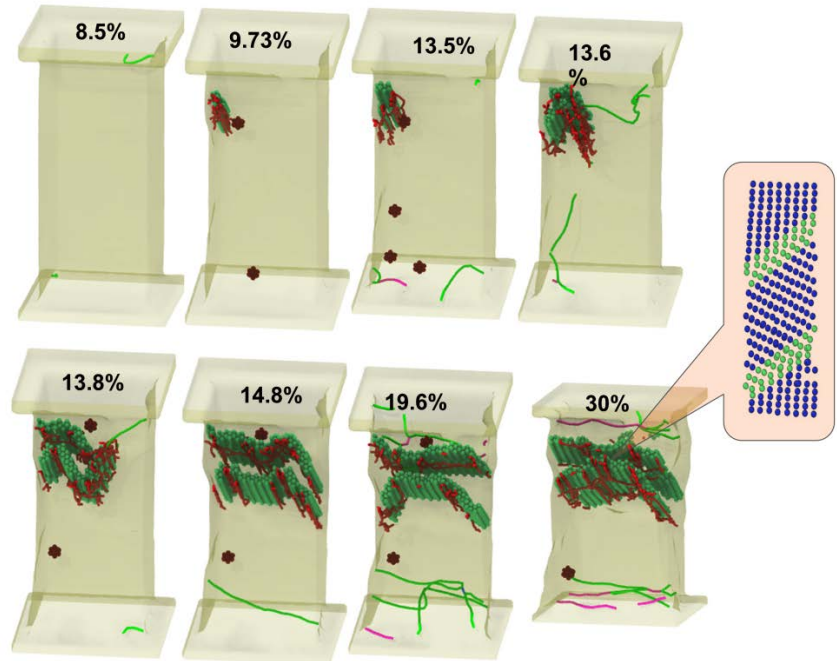

**Figure 8** Deformation of the nanopillar at 300 K. Dislocation storage can be seen to be smaller than that in Fig. 7. The inset shows part of the twin slab with displaced boundaries.

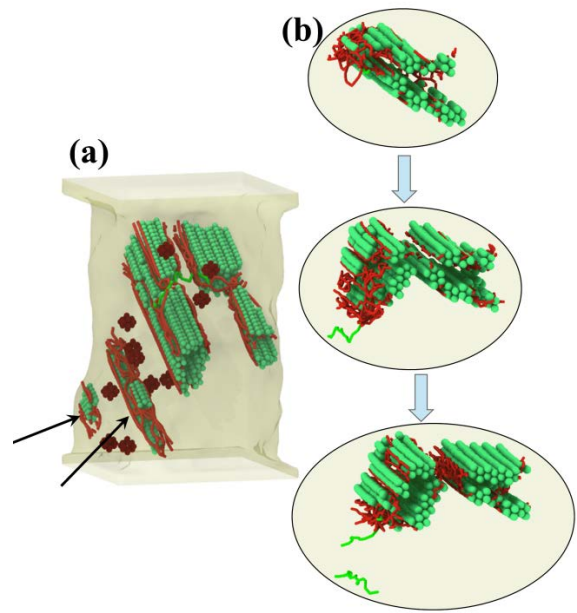

**Figure 9** (a) Four distinct twin faults of finite extents formed by the dissociation of four <111>/2-screws at 300 K. (b) Emission of two primary dislocations by the tip of a twin-fault. These emissary line defects further dissociate due to extrinsic pinning at point defects, thereby creating the two twin faults indicated by arrows in panel (a).

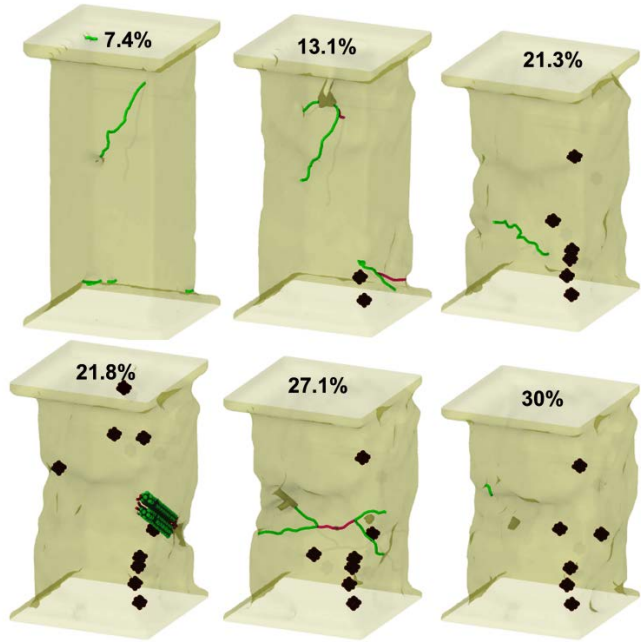

**Figure 10** Compressive deformation of the nanopillar at 500 K. Dislocation storage is almost absent at such high temperature.

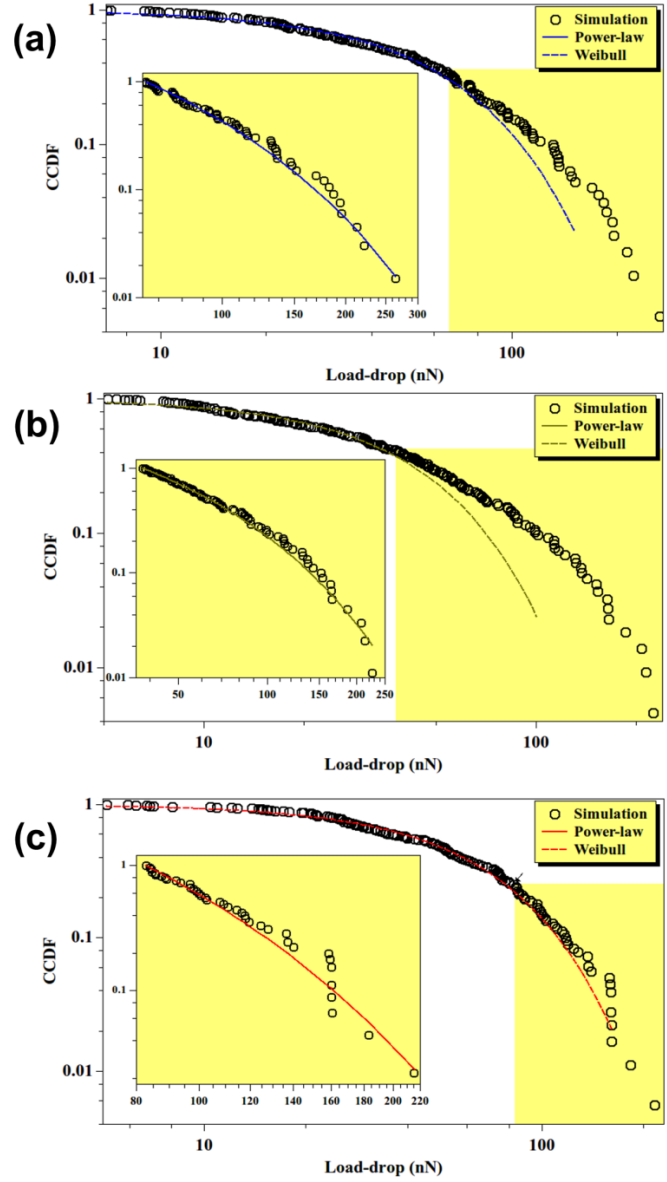

**Figure 11** Complementary cumulative probability distributions of the load-drops at (a) 50 K, (b) 300 K and (c) 500 K temperatures. In each plot, the yellow highlighted region covesr the regime of large load-drops with power-law distributions. CCDFs of the power-law regimes are separately plotted in the accompanying insets. Dotted lines show the fits corresponding to Weibull distributions. Power-law fits are given by solid lines.

**Table 1** Parameters corresponding to yield point and statistical distributions at the three studied temperatures. Refer to Sec. 3.4 for interpretation of the statistical parameters.

| Parameters | 50 K | 300 K | 500 K |
| --- | --- | --- | --- |
| Yield stress | 13.3 GPa | 9.6 GPa | 8.2 GPa |
| Yield strain | 11.8% | 8.8% | 7.3% |
| $\alpha$ | 0.42 | 0.37 | 0.43 |
| $\lambda$ | 58.7 nN | 38.3 nN | 62.3 nN |
| $\beta$ | 1 | 1 | 1 |
| $\mu$ | 64.9 nN | 71.7 nN | 44.8 nN |